\documentclass[fleqn,usenatbib]{mnras}

\usepackage{newtxtext,newtxmath}

\usepackage[T1]{fontenc}

\DeclareRobustCommand{\VAN}[3]{#2}
\let\VANthebibliography\thebibliography
\def\thebibliography{\DeclareRobustCommand{\VAN}[3]{##3}\VANthebibliography}

\usepackage{graphicx}	
\usepackage{amsmath}	
\usepackage{tabularx}

\title{Detection Statistics of Pulse Signals at Declinations from $+42^{\circ}$ to $+52^{\circ}$ at the Frequency 111 MHz}

\author[V. A. Samodurov et al.]{
V. A. Samodurov $^{a, b}$\thanks{E-mail: sam@prao.ru}
S. A. Tyul'bashev,$^{b}$
M. O. Toropov,$^{c}$
S. V. Logvinenko,$^{b}$
\\
$^{a}$ National Research University Higher School of Economics, Moscow, 109028 Russia,\\
Radiotelescopnaya 1a, Moscow reg., Pushchino, 142290, Russia \\
$^{b}$ Pushchino Radio Astronomy Observatory, Astro Space Center, Lebedev Physical Institute, Russian Academy of Sciences, Pushchino, 142290 Russia\\  
$^{c}$ LLC TEK Inform, Moscow, 117246 Russia \\
}

\date{December 12, 2021}

\pubyear{December 27, 2021}

\begin{document}
\label{firstpage}
\pagerange{\pageref{firstpage}--\pageref{lastpage}}
\maketitle

\begin{abstract}
A search for pulse signals was carried out in a new sky area included in the monitoring program for the search for pulsars and transients. Processing of several months data recorded in six frequency channels with a total bandwidth of 2.5 MHz showed that, on average, 4 pulses per hour are observed in each of the 24 connected stationary beams. Of these pulses, $18.3\%$ are similar to those of pulsars. They are visible in one or two neighboring beams and have a pronounced dispersion shift, that is, they are recorded first at a high and then at a low frequency, which indicates the possible passage of the signal through the interstellar medium. Almost $68\%$ of such detected pulses belong to six known pulsars with dispersion measures from 9 to 141 $~pc/cm^3$, and almost all of the remaining pulses are either noise of an unknown nature or artifacts of the proposed pulse separation technique. An additional study of the selected array of 3650 obvious pulsar pulses revealed 13 pulses belonging to four rotating radio transients (RRATs). Their dispersion measures are in the range of 17–51 $~pc/cm^3$. A search for regular (periodic) RRAT emission was carried out using power spectra summed over 121 days. Periodic radiation was not detected, but for two RRATs, upper estimates of the periods were obtained from measurements of the time intervals between pulses. The upper estimates of the integrated flux density of the detected RRATs are in the range 2–4 mJy at the frequency 111 MHz. 
\end{abstract}




\section{Introduction}
Initially, pulsars were discovered as sources of pulsed dispersed radiation (\citeauthor{Hewish1968}, \citeyear{Hewish1968}). The observed pulses were located at the same time intervals (periods) from each other. The pulses could be summed up and, if the period is known with sufficient accuracy, the signal-to-noise ratio ($S/N$) could be improved. The obvious logic tells us that in order to detect weak pulsars, we need to add many periods, that is, other things being equal, increase the total observation time. In addition, due to the dispersion of the signal in the interstellar medium, observations must be carried out in many frequency channels, and due to the short pulse duration of the pulsar, one point readout time from hundreds of microseconds to milliseconds should be used. The fulfillment of these conditions led to the accumulation of large volumes of raw data that had to be processed using the weak, at that time, capabilities of computer technology.

Almost immediately after the discovery of pulsars, it was proposed to search for them using fast algorithms to detect periodic signals. The search in the frequency domain could be performed using the Fast Fourier Transform (FFT), and the search in the time domain could be made using periodograms (\citeauthor{Lovelace1969} (\citeyear{Lovelace1969}), \citeauthor{Staelin1969} (\citeyear{Staelin1969}), \citeauthor{Burns1969} (\citeyear{Burns1969})). Obvious steps to speed up the processing of observations led to the fact that for decades the method of searching for pulsars by their individual pulses was left aside.

In the early history of pulsar astronomy, the study of individual pulses was of little interest, since the shape of individual pulses and their intensity vary widely. A more stable structure is the average profiles, in which hundreds and thousands of pulses are summed. The shape of the average profiles reflects the geometric features of the pulsar magnetosphere structure. Nevertheless, the study of the pulses from the strongest pulsars has shown the existence of pulse microstructure, subpulse drift, giant pulses, nullings, and mode switching (see the handbook on pulsar astronomy (\citeauthor{Lorimer2004}, \citeyear{Lorimer2004}) and references therein).

In 2003, a paper \citeauthor{Cordes2003} (\citeyear{Cordes2003}) was published, in which the optimal search for pulse signals was considered considering their scattering and scintillations by the interstellar medium. In 2006, rotating radio transients (RRATs) were discovered, that is, pulsars, which can have from tens of seconds to hours between detected pulses (\citeauthor{McLaughlin2006}, \citeyear{McLaughlin2006}). Periodic radiation is often not detected between successive strong pulses. In 2007, fast radio bursts (FRBs) were discovered, which are pulse signals coming from outside the Milky Way (\citeauthor{Lorimer2007}, \citeyear{Lorimer2007}).

The discovery of extraterrestrial pulse signals that cannot be detected using standard periodic emission searches has brought the search for dispersed pulse signals back to life as an additional search method for conventional searches for pulsars. The search for these dispersed signals is associated with a number of difficulties. First, to search for and study individual pulses, radio telescopes are needed that have a high instantaneous sensitivity, which makes it possible to find a single pulse with an acceptable ($S/N$). For comparison, consider the search for an ordinary pulsar with a period $P_0=1$~s. A ten-minute recording of such a pulsar is enough to magnify ($S/N$) by almost 25 times. That is, a pulsar from individual pulses of which even a trace is not visible in the raw record can be studied without any problems with a radio telescope with average characteristics. Secondly, when searching for new transients, we $a priori$ do not know either the pulse dispersion measure or the width of its profile, which increases the number of searches during the search. At the same time, in raw data, the width of the noise track can change due to changes in the background temperature, while changing the standard deviations of the noise. This leads to the need for constant monitoring of noise at local points. In addition to this, at low frequencies (meter wavelength range), a scintillation (compact) source passing through the line of sight simultaneously with a pulsed signal can lead to track broadening. The scintillations of a compact radio source by the interplanetary plasma broaden the noise track. If a scintillating source and a dispersed pulse enter the antenna beam at the same time, the observation conditions may deteriorate. In addition, noise is regularly detected in the recordings, the pulses of which may have signatures of dispersion. Therefore, it is difficult to develop a search system that will unambiguously separate dispersed pulses of an extraterrestrial nature from interference. The situation of separating real signals from noise in the search for classical pulsars looks much better. The recurrence of signals in sidereal time, the same periods, the same dispersion measure, the similarity of the average profiles for different days, and the possibility of signal accumulation over different days unequivocally testify in favor of the discovery of a new pulsar.

In the present paper, within the framework of the organized survey Pushchino Multibeam Pulsar Search (PUMPS, \citeauthor{Tyulbashev2016} (\citeyear{Tyulbashev2016}), \citeauthor{Tyulbashev2018a} (\citeyear{Tyulbashev2018a}), \citeauthor{Tyulbashev2021} (\citeyear{Tyulbashev2021})), we study the types of short-duration signals found in raw data. We consider the statistics of detections for pulse radiation sources in a new sky area included in the monitoring program for the search for pulsars and the identification of these sources with interference and real signals of extraterrestrial origin.

\section{OBSERVATIONS AND DATA PROCESSING}

Observations were carried out with the Meridian type telescope Large Phased Array (LPA) antenna of the Lebedev Physical Institute (LPI). The receiving center frequency was 110.25 MHz and the usable bandwidth was 2.5 MHz. After a major reconstruction of the antenna, it became possible in principle to make four independent radio telescopes on the basis of one antenna field (\citeauthor{Tyulbashev2016} (\citeyear{Tyulbashev2016}), \citeauthor{Shishov2016} (\citeyear{Shishov2016})). In this paper, we consider a radio telescope (LPA3) used for monitoring observations. A feature of the LPA3 antenna is that it has a system of beams fixed in directions. A total of 128 beams were implemented, covering declinations in the meridian plane of the declinations from $-9^{\circ}$ to $+55^{\circ}$. In 2013–2014, 96 beams were connected to the recorders, covering declinations from $-9^{\circ}$ to $+42^{\circ}$. At the end of 2020, a new recorder was created, to which 24 more beams were connected in test mode. Currently, the new recorder is used to check individual connected beams, frequency channels, amplifier operation, the amount of interference, the quality of observations in general, and so on.

Amplifiers of the first floor are located at the output of the dipole lines of the LPA LPI. A known temperature signal (calibration signal; calibration step) can be applied to the input of these amplifiers, while turning off the dipole lines themselves. This signal passes through all antenna paths and all amplifiers located along the way. The calibration signal is recorded in the form OFF–ON–OFF, where the OFF mode means the  absence of a calibration signal when all intermediate amplifiers and antenna (dipole lines) are turned off. In this case, the noise in the antenna paths (cables) corresponding to the ambient temperature is prescribed. The ON mode is the inclusion of the calibration signal when the dipole lines are turned off. Since the recording of the calibration step takes place in all frequency channels, it is possible to equalize the gain in each frequency channel independently (for more details, see \citeauthor{Tyulbashev2019} (\citeyear{Tyulbashev2019})).

The operating mode of the new recorder is the same as that of the recorders used for operation in 96 beams (\citeauthor{Tyulbashev2016}, \citeyear{Tyulbashev2016}). At the beginning, the full band of observations is digitized. In the recorder, the input data stream is converted into data with low (0.1 s; 6 frequency channels) and high (12.5 ms; 32 frequency channels) time-frequency resolution, after which the data in both formats are written to hard disks. The beginning and end of the recording for each file with raw data coincides with the beginning and end of the next hour in UT.

Previously, data with a low frequency-time resolution were used in the Space Weather project (\citeauthor{Shishov2016}, \citeyear{Shishov2016}) and also in the search for pulsars (\citeauthor{Tyulbashev2016}, \citeyear{Tyulbashev2016}). For the same purposes, data from the new receiver will also be used. Since data are recorded in both formats simultaneously, it is possible to assess the quality of observations as a whole by processing only data with a low frequency- time resolution. The volume of these data is approximately 35 times smaller than that of data with high frequency-time resolution. In this paper, we consider the quality of data in 24 beams connected to the new recorder, covering declinations from $+42^{\circ}$ to $+52^{\circ}$.

Since the LPA is a meridian instrument, the source under study can be observed only during its passage through the meridian. The transit time is approximately 3.5 minutes at declination $\delta =0^{\circ}$ along the beam width at half-power points. The beam shape is subject to dependence $[sin(x)/x]^2$, and therefore, when estimating the flux density, it is necessary to make corrections that consider the characteristics of the antenna array.

Before processing the observations, the gain in the frequency channels is equalized using a calibration step. The data quality and noise environment are then evaluated over ten second time slices using standard statistical methods. For each time interval, the minimum and maximum intensity values in arbitrary units are estimated before and after the removal of pulse noise, the median value of the intensity in the studied segment, and standard deviations. The date and hour of observations are stored in Moscow time, as well as the beginning of the ten-second segment under study in sidereal time. The comparison of standard deviations from day to day and from one segment to another during the day allows us to control the level of interference both on a given day and on any selected time interval.

\section{RESULTS AND DISCUSSION}

\begin{figure*}
\begin{center}
	\includegraphics[width=1.0\textwidth]{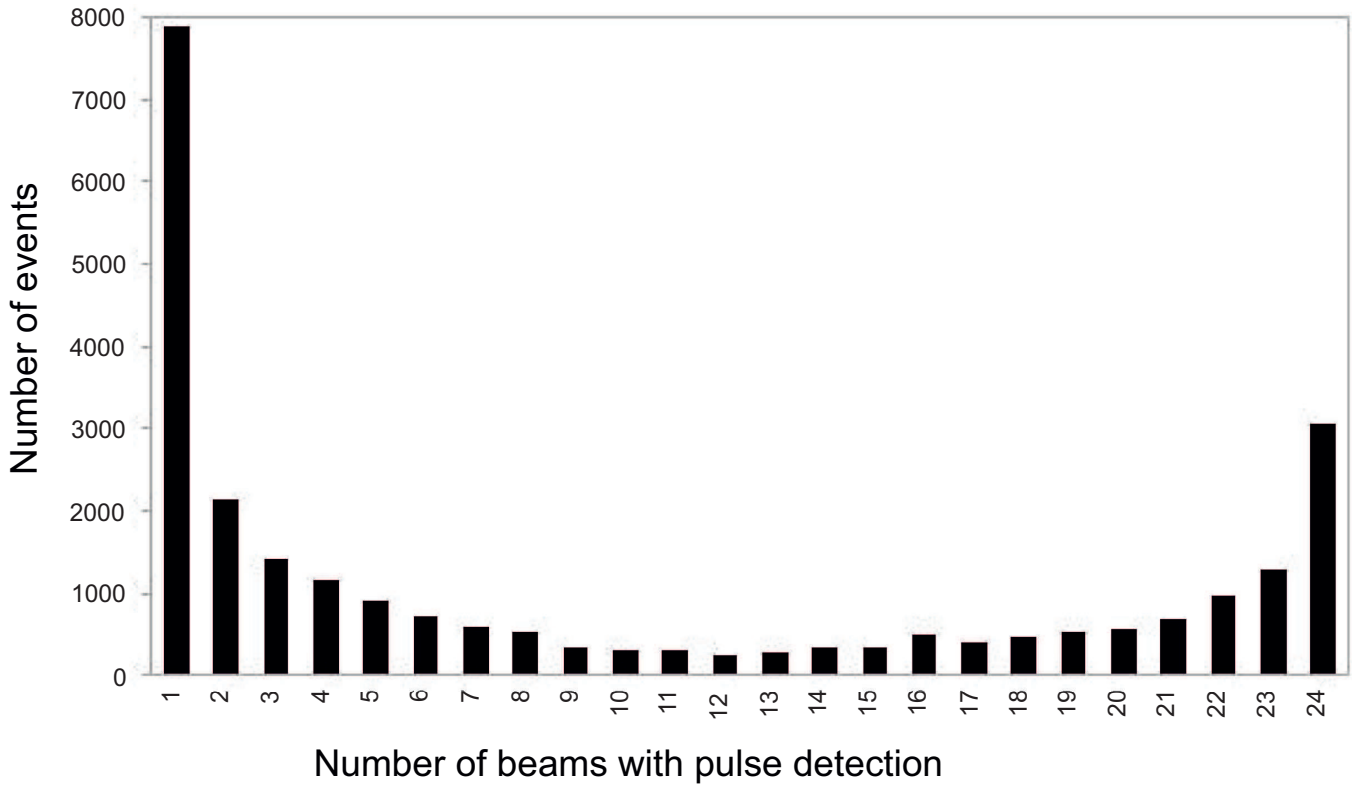}
    \caption{Pulse distribution histogram: the vertical axis shows the number of detected pulses, and the horizontal axis shows the number of beams in which the pulse was detected.}
    \label{fig:fig1_rrat}
\end{center}
\end{figure*}

To check the quality of observations and to understand the nature of pulse noise, a semi-annual interval of December 16, 2020–August 24, 2021 was chosen. Hourly data files were calibrated with a calibration step and then divided into 10 second segments for analysis. In each frequency channel, an independent search was carried out for pulse signals with $S/N >5$ ($S/N =A/\sigma_{noise}$, where $A$ is the signal amplitude after subtracting the baseline (background signal), and $\sigma_{noise}$ is the standard deviations at a 10-second interval. If on the studied segment for the found pulse $S/N$ was more than 5 in at least three frequency channels out of six, then this segment was studied separately. A similar technique, but with different intervals (5 s) and using database mechanisms, has already been applied when working with data for 2012–2013 (\citeauthor{Samodurov2017}, \citeyear{Samodurov2017}). An earlier work recorded 83 086 pulses at declinations from $+3^{\circ}$ to $+42^{\circ}$. In particular, two new RRAT pulsars were discovered in this paper for the first time.

For each pulse signal found in the frequency channel, the time when its maximum was observed was recorded. If the pulse maximum at a lower frequency is observed later than at a higher frequency, this indicates a possible dispersion of the signal caused by the interstellar medium. Such candidates were additionally checked. Information was also recorded about pulses that had the opposite (“anti-pulsar”) behavior, that is, a pulse at a low frequency arrived earlier than at a higher frequency. For each of the 10-second data segments, the stored pulse passports included information about the beam number, Modified Julian Date (MJD), time (UT), sidereal time, $S/N$ observed in the frequency channels, the expected dispersion measure ($DM$), the label of a pulsar candidate, or expected interference, information was recorded on how many beams an pulse was observed, lists of numbers of beams with similar pulses. In particular, information about how many beams simultaneously observed a pulse makes it possible to detect interference. Since the noise is not directional, they are recorded in many or even in all 24 investigated beams.

Over the entire observation interval, 3.2 million pulse events were detected in 24 beams. The verification showed that after April 15, 2021, the vast majority of events were observed in two spatial beams, which indicates technical problems. The reasons for the decrease in the quality of observations are being clarified. We attribute such a sharp increase in the number of events to the new frequency filters delivered for testing, which do not cut out frequency noise well.

For further analysis, data from December 16, 2020 to April 15, 2021 were selected, i.e., 121 days of observations. Considering equipment shutdowns for scheduled maintenance and hardware failures, the final analysis was carried out for 2739 hourly files or 114.125 days.

247 881 pulses were detected in the indicated interval. The typical number of detected pulses is 10– 11 thousands in each of 24 beams, i.e., an average of about 4 events per hour per beam. In this case, the pulses often arrived at the same time moment (with an accuracy of fractions of a second), in different beams of the LPA diagram. We combined such pulses into one event. As a result, these pulses (247 881) are combined by us into 26 011 events. For the found dispersed pulses, the distribution of simultaneous events over the beams is shown in Fig.~\ref{fig:fig1_rrat}. The histogram shows a two-hump distribution of events, where the maxima fall on events visible only in one beam and on events observed in all 24 beams. The left hump should contain all the events associated with the detection of real pulses, plus, possibly, part of the noise. The right hump should contain all interference cases (technogenic noise) plus, perhaps, the most powerful pulsar pulses can be observed in many beams, appearing in the side lobes of the LPA LPI (see, for example, \citeauthor{Tyulbashev2021a} (\citeyear{Tyulbashev2021a})).

Fig.~\ref{fig:fig1_rrat}, in total, shows 26 011 events, i.e., 9.5 events (simultaneous pulses in different beams) per hour. Of these, 4750, that is, about $18.3\%$ of the events are similar to pulsar pulses, since they were visible in no more than 2 beams and showed a noticeable dispersion shift. Combining simultaneous pulse observations in two beams as one event, we get 4750 unique events. All these pulses were separated into a separate text file for a more detailed analysis.

The finished text file, consisting of lines with descriptions of pulses, was analyzed by a program that allows checking clusters of pulses for different sets of features. When running the program, the following features were used: right ascension coordinate, declination coordinate, the number of recorded pulses over the entire observation time, and the average dispersion measure in a set of pulses with close coordinates. The size of the analyzed time interval was two minutes in right ascension for the beam. That is, the beam is divided into 30 (intervals per hour) $\times$ 24 (hours), a total of 720 segments. For 24 beams, respectively, 17 280 unique time intervals are created, within which pulses are accumulated for their further analysis.

In total, the processing program identified 470 time segments among 17 280 possible segments in which at least one pulsar (dispersed) pulse was detected. Each of the segments was checked according to the ATNF catalog (($https://www.atnf.csiro.au/research/pulsar/psrcat/$, \citeauthor{Manchester2005} (\citeyear{Manchester2005})). All doubtful cases were checked visually and discarded from the array of analyzed pulses. In particular, since more than one pulse event could occur within the automatically analyzed 10-s data segments (for example, two successive pulsar pulses could pass), processing artifacts occurred when, at one frequency, the pulse time was taken from one such event within 10 s, and at the second frequency, the pulse time was taken another event. Such cases lead to falsely detected events with anomalously high or, conversely, negative dispersion delays (up to DM=$\pm 1900$ pc/cm$^3$). All such cases were removed from the sample of pulses and were not analyzed. As a result, out of 4750 pulsarlike events (in no more than two LPA beams and having a dispersion shift), 3650 pulses were left visually similar to real pulsars, that is, $76.8\%$ of the pulses automatically identified by the program and marked as candidates for pulsar pulses.

\begin{figure*}
\begin{center}
	\includegraphics[width=1.0\textwidth]{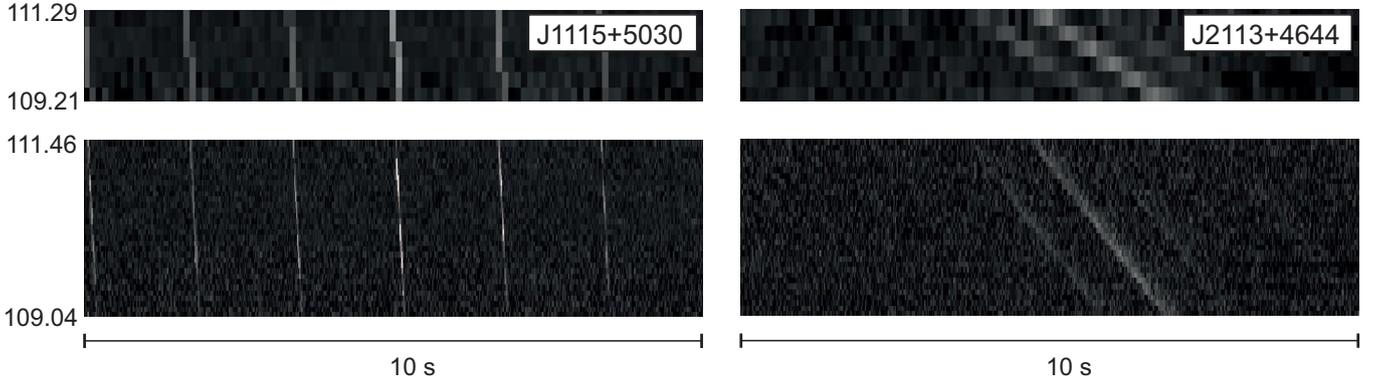}
    \caption{The left and right parts of the figure show the dynamic spectra of the pulsars J1115+5030 ($P_0=1.6564$~s; $DM=9.18$~pc/cm$^3$) and J2113+4644 ($P_0=1.0146$~s; $DM=141.26$~pc/cm$^3$), respectively. The horizontal scale shows the time. Each dynamic spectrum is 10 s of the same record in 6 (top) and 32 (bottom) frequency channels. The vertical scale shows the frequency channels. Two parallel lines of light squares on the upper right dynamic spectrum clearly visualize all six frequency channels, where the upper pixel on the vertical axis of the dynamic spectrum corresponds to the channel center frequency 111.29 MHz, the lower pixel corresponds to a frequency 109.21 MHz with a step of 415 kHz. In the bottom panels, the top of the 32 channels corresponds to the channel center frequency 111.461 MHz and the bottom channel corresponds to 109.039 MHz with a step of 78 kHz.}
    \label{fig:fig2_rrat}
\end{center}
\end{figure*}

The first column of the histogram (Fig.~\ref{fig:fig1_rrat}) contains almost 8000 points associated with pulses with or without dispersion delays. About half of them turned out to be associated with pulses of known pulsars: B0011+47 (J0014+4746), B1112+50 (J1115+5030), J1955+5059, B2021+51 (J2022+5154), B2111+46 (J2113+4644), and B2217+47 (J2219+4754). As an illustration, Fig.~\ref{fig:fig2_rrat} shows the dynamic spectra of the pulsars J1115+5030 and J2113+4644.

Obviously, in 6-channel data, the sensitivity is lower than in 32-channel data due to the dispersion spreading (smoothing) of the signal within one frequency channel and due to the fact that the sampling time in the recorded data is greater than the pulse width. Nevertheless, despite this decrease in sensitivity, all pulses from the pulsar J1115+5030  are visible without gaps in the dynamic spectrum. Note also that pulses from the pulsar J2113+4644 were found in the blind search. A large dispersion measure and, consequently, a strong dispersion smoothing of the pulse in the frequency channel, as well as scattering in the interstellar medium, should greatly reduce the observed peak flux density. Nevertheless, two successive pulses from J2113+4644 are clearly visible in Fig.~\ref{fig:fig2_rrat}. Table~\ref{tab:tab1} shows the statistics of detection of pulsar pulses for 121 processed days, or for 114.125 days, taking into account prevention and observation failures. The first column gives the name of the pulsar, columns 2 and 3 show $P_0$ and $DM$ from ATNF, columns 4 and 5 present the total number of registered pulses ($N1$) and for how many days they were detected ($N2$).

\begin{table}
\centering
\caption{Statistics of pulse detections of known pulsars.}
\label{tab:tab1}
\begin{tabular}{ccccc}
\hline
Name & $P_{0},s$ & DM (pc/cm$^3$) & $N_1$, pulses & $N_2$, days \\
\hline
J0014+4746 & 1.2406 & 30.4   & 30   & 21\\
J1115+5030 & 1.6564 & 9.18   & 1245 & 108\\
J1955+5059 & 0.5189 & 31.9   & 6    &  6\\
J2022+5154 & 0.5291 & 22.5   & 102  &   58\\
J2113+4644 & 1.0146 & 141.2  & 63   & 44\\
J2219+4754 & 0.5384 & 43.4   & 2191 & 107\\
\hline
\multicolumn{5}{p{8cm}}{The pulsars J2113+4644 and J2219+4754 are also observed in the side \newline lobes of the LPA LPI beam.}\\
\hline
\end{tabular}
\label{tab:tab1}
\end{table}

\begin{table*}
	\centering
	\caption{Characteristics of discovered pulsars.}
	\label{tab:tab2}
	\begin{tabular}{cccccccccc} 
		\hline
Name & $\alpha_{2000}$ (h,m,s) & $\delta_{2000}$  ($^{\circ} \, ^{\prime} $) & b ($^{\circ} \, ^{\prime}$) & $P_0$ s & $DM$ pc/cm$^3$ & $W_{0.5}$ (ms) & $S_{peak1,2}$ (Jy) & $N_1/N_2$  & $n$, 1/h\\
		\hline
J0939+45 & 09 39 31 & 45 15 & 48 00  & -     &  17.5$\pm$1 & 23.5 (49.5) & 16.4, 4.9  & 5(6)/5 & 1\\
J1218+47 & 12 18 56 & 47 14 & 68 54  & -     & 19.4$\pm$1  & 21.0 (30.4) &  31.3, 9.9 & 4(4)/4 & 0.67\\
J1929+42 & 19 29 11 & 42 40 & 11 38  & 3.6375& 51.25$\pm$2 & 36.9 (52.5) & 11.9, 8.3  & 3(5)/3 & 0.83\\
J2214+45 & 22 14 01 & 45 25 & -9  06 & 2.725 & 19.15$\pm$1 & 18.0 (20.8) &  12.7, 4.3 & 1(3)/1 & 0.5\\
		\hline
	\end{tabular}
	\label{tab:tab2}
\end{table*}

In addition to known pulsars, pulses belonging to 4 RRATs were found in the records. For these transients, from 1 to 5 pulses were initially found in the 6-channel data. Recall that in the initial search, sources were selected if the pulse was detected in at least half (in the case of 6-channel data, three) frequency channels with $S/N > 5$. The ATNF catalog makes it possible to extract the median value of the pulse width of second pulsars, which falls by about 30–40 ms. Therefore, for a typical second pulsar, the loss of sensitivity in 6-channel data due to too long a sampling time for one point will be equal to $(100/35)^{1/2}=1.7$ times. If the weakest pulse was observed with $S/N = 5$ in all six channels, then when processing 32-channel data, the weakest pulses will have $S/N = 5\times 6^{0.5} \times 1.7 \approx 21$. This means that in the 32-channel data, by sorting through the dispersion measures close to the expected one, we can try to find new pulses that were missed in the 6-channel data.

We processed the 32-channel data for the days when the new RRAT pulses were recorded in order to try to detect possible weak pulses lost during processing of the 6-channel data and, if possible, to obtain upper estimates of the RRAT periods. After reprocessing the data for days in which new RRAT pulses were previously detected, the number of detected pulses increased. It became possible to make an upper estimate of the period for RRATs that have several pulses during a given session. The period estimate can be obtained as the greatest common divisor for all time intervals between individual pulses. The present period can be an integer number of times less. Using 32-channels data on strong pulses, a re-evaluation of $DM$ for the studied RRATs was made. To do this, the shifts in the frequency channels were enumerated with a step of unity in $DM$ and in this enumeration it was considered that the absolute maximum in the pulse summed over all frequency channels corresponds to the correct $DM$.

In Table~\ref{tab:tab2} we provide information on the found RRATs. The first column contains the name of the transient. In the second—his right ascension. Right ascension was defined as the average of the right ascensions of all pulses. Since the total number of pulses found is small, the error is defined as $\pm 1.5^m$, that is, by half the power of the LPA radiation pattern. The third column gives the declinations. The found pulses are visible in one beam and are not visible in the beams above and below; therefore, the declination coordinate was determined as the beam declination, and the accuracy was determined as half the declination distance between the beams ($\pm 15^{\prime}$). The fourth column shows the galactic latitude. Columns 5–8 give estimates of the period, dispersion measure, halfwidth of the profile of the narrowest and, in parentheses, the widest detected pulse, peak flux densities of the strongest ($S_{peak1}$) and weakest ($S_{peak2}$) found pulses. The peak flux density in pulse was estimated based on the observed of the pulse and the expected sensitivity of the LPA LPI in the direction of the transient. The given estimates are the lower estimates of the flux density. The exact coordinate of the transient, neither in right ascension nor in declination, is known. Therefore, it is impossible to make corrections that consider the possible impact of the pulse on the edge of the LPA radiation pattern and the possible mismatch between the coordinates of the LPA beam and the declination coordinates of the transient. As a result, the estimate of the peak flux density can be underestimated by a factor of 1.5–2. The ninth column indicates how many pulses were detected ($N_1$) from the data with low and (in brackets) with high time-frequency resolution. Separated by the sign "/", information is given on how many days the pulses were detected ($N_2$). The tenth column shows the frequency of occurrence of pulses ($n$) that have $S/N >10$ in 32-channel data per hour of observations. When obtaining the estimate of $n$, it was assumed that most likely the pulses appear in the central part of the LPA3 beam pattern, which is approximately equal to 3 min at half power. For 120 days, $(120\times 3)/60=6$ hours of observations have accumulated in the direction of each transient.

When identifying the found RRATs in the ATNF catalog, it turned out that the source J1929+42 we detected probably coincides with the source discovered from individual pulses with the CHIME radio telescope (\citeauthor{Good2020}, \citeyear{Good2020}). In this paper, 63 pulses were detected, but only 4 of them were discovered at the pulsar facility, and the authors of the paper failed to obtain an estimate of the period. For the rotating radio transient found in CHIME $\alpha_{2000}=19^h31^m\pm 7^m$, $\delta_{2000}=42^o30^\prime \pm 5^\prime$, $DM=50.9$ pc/cm$^3$ and $W_{0.5}=32$~ms, the peak flux density was observed for four pulses at 600 MHz in the range from 25 to 150 mJy. On average, 8 pulses per hour were recorded. The spectral index $\alpha =2-2.5 (S\sim \nu^\alpha$) is consistent with flux density estimates at 111 and 600 MHz. Thus, our data from Table 2 for the source J1929+42 for the coordinates, pulse halfwidth, and flux density for LPA and the data for CHIME indicate an unambiguous identification of the sources.

\begin{figure*}
\begin{center}
	\includegraphics[width=1.0\textwidth]{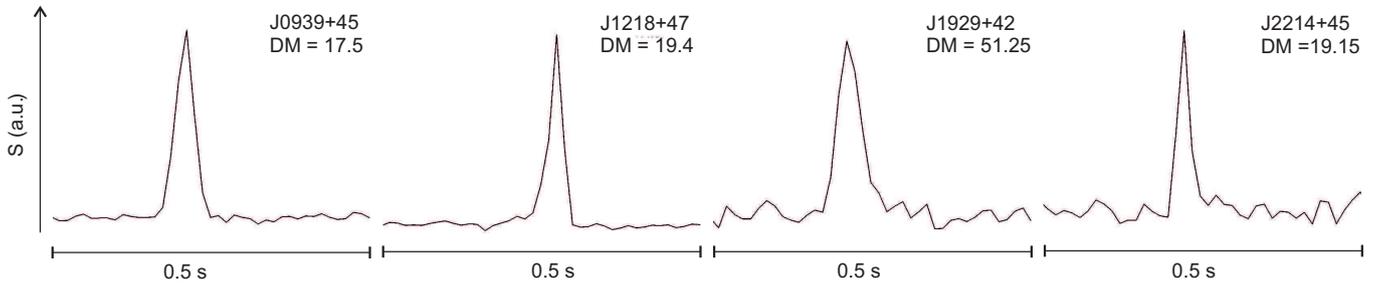}
    \caption{Profiles of the strongest pulses of the detected transients. Note that RRAT J0939+45 exhibits both single and double pulse profiles. The horizontal axis shows the time scale, and the vertical axis indicates the flux density in arbitrary units (a.u.).}
    \label{fig:fig3_rrat}
\end{center}
\end{figure*}

In addition to sporadic pulsed emission, many RRATs discovered on individual pulses at high frequencies also exhibit regular (periodic) emission in the meter wave range (see, for example, \citeauthor{Losovsky2014} (\citeyear{Losovsky2014})). We tried to find the periodic emission for the open RRATs using the power spectra and their subsequent summation over all observation days. This method makes it possible to raise the sensitivity approximately to the root of the number of stacked power spectra (\citeauthor{Tyulbashev2017} (\citeyear{Tyulbashev2017}), \citeauthor{Tyulbashev2020} (\citeyear{Tyulbashev2020})). The expected increase in sensitivity in the search for regular radiation was approximately 10 times. Harmonics with $S/N >5$, indicating a periodic signal, were not found in the power spectra. Assuming the worst case of the location of the pulsar between the LPA beams and taking into account the sensitivity of observations in the plane and out of the plane of the Galaxy (\citeauthor{Tyulbashev2016}, \citeyear{Tyulbashev2016}), we exclude the regular (pulsar) emission of the investigated RRATs at the level: J0941+45 (($S_{int}<2$ mJy), J1218+47 (($S_{int}<2$ mJy), J1929+42 (($S_{int}<4$ mJy), and J2214+45 (($S_{int}<4$ mJy). Samples of the pulse profiles of these RRATs are shown in Fig.~\ref{fig:fig3_rrat}.

\section{CONCLUSIONS}

The monitoring program for observations based on data recorded with a low time-and-frequency resolution showed high efficiency, making it possible to promptly respond to changes in both external observation conditions and internal causes responsible for the deterioration of observations. A relatively small number of recorded pulses of a noise nature per one hour of observations shows that, in general, the quality of observations is high.

Over a 121-day period of observations, six-channel data with a channel width of 415 kHz and a sampling rate of one point of 0.1 s detected more than 4750 events similar to pulses of pulsars. For each of these events, a pulse at a high frequency arrives earlier than at a low frequency, and is recorded in no more than two neighboring beams. Among the indicated events, 3650, i.e., $76.8\%$ turned out to be real pulsar events. The remaining pulsar-like events turned out to be associated either with rare data failures or with artifacts of the proposed data processing algorithm.

The blind search found 6 known pulsars with 6 to 2191 pulses in the low time-frequency data over 121 observation days. In addition to the known pulsars, 4 RRATs have been discovered. One of the detected RRATs apparently coincides with a source previously discovered with the CHIME radio telescope. \newline 

{ACKNOWLEDGMENTS}

The authors thank V.V. Oreshko for the prompt activation of new beams and the antenna group for providing observations, as well as L.B. Potapova for help in preparing the paper.

\bsp	
\label{lastpage}
\end{document}